# Portfolio Optimization with Robust Covariance and Conditional Value-at-Risk Constraints

**Qiqin Zhou**
qz247@cornell.edu
Cornell University

**Abstract**

The measure of portfolio risk is an important input of the Markowitz framework. In this study, we explored various methods to obtain a robust covariance estimators that are less susceptible to financial data noise. We evaluated the performance of large-cap portfolio using various forms of Ledoit Shrinkage Covariance and Robust Gerber Covariance matrix during the period of 2012 to 2022. Out-of-sample performance indicates that robust covariance estimators can outperform the market capitalization-weighted benchmark portfolio, particularly during bull markets. The Gerber covariance with Mean-Absolute-Deviation (MAD) emerged as the top performer. However, robust estimators do not manage tail risk well under extreme market conditions, for example, Covid-19 period. When we aim to control for tail risk, we should add constraint on Conditional Value-at-Risk (CVaR) to make more conservative decision on risk exposure. Additionally, we incorporated unsupervised clustering algorithm K-means to the optimization algorithm (i.e. Nested Clustering Optimization, NCO). It not only helps mitigate numerical instability of the optimization algorithm, but also contributes to lower drawdown as well.

## 1  INTRODUCTION & LITERATURE REVIEW

Under the Markowitz framework, a portfolio relies heavily on the availability of the covariance matrix. Estimating covariance is considered more practical than predicting asset returns. Therefore, many researchers focus on constructing minimum-variance portfolios that rely solely on the estimation of the covariance matrix. However, resulting portfolios can still perform poorly when tested out of sample.

The traditional estimator is the sample covariance matrix [1] (Jobson and Korkie, 1980), but it is known for its non-robustness and instability. When the sample covariance matrix is used in an optimization problem, the result is prone to high estimation errors due to noisy inputs and the signal structure. Existing literature has attempted to resolve this problem in three ways. The first approach is to introduce strong constraints on weights, such as short-sales constraints [3] (Frost and Savarino, 1988). Secondly, Ledoit and Wolf (2003) suggested shrinking the covariance matrix towards a pre-specified, more stable structure [4]. The third method is to introduce more robust covariance estimators that rely less on product-moment. For example, the Minimum Covariance Determinant (MCD) estimator is a robust estimator that can be used to estimate the covariance matrix of highly contaminated datasets [6] (Rousseeuw, 1984). Additionally, Marcos López de Prado (2019) proposed reducing signal instability by clustering the covariance matrix into subgroups using K-means or Hierarchical Clustering [10].

Another reason why the traditional Markowitz framework can fail in reality is that it pays little attention to controlling extreme risks. After the financial crisis, Value-at-Risk (VaR) became one of the most common risk measures used in finance. Rockafellar and Uryasev (2000) proposed a more coherent and sub-additive percentile risk measure: Conditional Value-at-Risk (CVaR), which calculates the expected loss given that the loss exceeds the VaR threshold [12]. Alexander and Baptista (2004) suggested that a CVaR constraint is more effective than a VaR constraint as a tool to control risk for slightly risk-averse agents [13].

The paper is organized as follows: we introduce the theoretical framework of covariance estimators and optimization problem formulation in Sections 2 and 3, respectively. We discuss data preparation and the rebalancing procedure in Section 4. We implement different covariance estimators within the framework of minimum variance optimization in Section 5. In Section 5.1, we compare the performance of sample covariance, Ledoit shrinkage covariance, and robust Gerber covariance on minimum variance portfolios. In addition to these covariance estimators, we also explore whether the implementation of Nested Clustering Optimization can help reduce instability caused by signal in Section 5.2. In Section 5.3, we add CVaR constraints to the minimum variance optimization to control for extreme risk. Specifically, we test and compare portfolios with one CVaR constraint and two CVaR constraints.

## 2  MODELS OF ESTIMATING COVARIANCE

In this section, we delve into the theoretical framework of covariance matrix estimation. We will explore four different covariance estimators: exponentially-weighted sample covariance, Ledoit-Wolf shrinkage covariance, Gerber robust co-movement measure, and Nested Cluster Optimization. The first three methods aim to construct more accurate and robust statistics for use in optimization, while the last method focuses on enhancing optimization results by clustering the correlation matrix to mitigate the spread of signal instability.

**Motivation**  Suppose we want to construct a minimum variance portfolio using traditional Markowitz framework. We have $N$ assets, whose variance is represented by a covariance matrix $V$. We want to compute the optimal weight vector $\omega^*$ of the following minimum variance optimization problem:

$$\min_{\omega} \frac{1}{2}\omega' V \omega$$
$$\text{s.t.: } \omega' e = 1,$$

where $e$ is vector of ones of size N

We can derive a closed-form optimal solution:

$$\omega^* = \frac{V^{-1} e}{e' V^{-1} e}$$

When we use $\hat{V}$ to estimate $V$, the solution can be unstable, which means a small change of the inputs will cause great change in $\widehat{\omega}^*$. This is the reason why portfolios using estimated covariance always perform worse out-of sample. For constructing a better portfolio, we want to study how we can mitigate the problem of instability.

### 2.1  Exponentially Weighted Sample Covariance

The first method is exponentially weighted sample covariance. It is an extension on traditional sample covariance matrix. Consider we have $N$ assets and $T$ observations (historical samples).

**Sample Mean Return** We first define the sample mean vector $\bar{\mathbf{x}}$ as a column vector whose $j$-th element $\bar{x}_j$ is the average value of the $N$ return observations of the $f^{\text{th}}$ variable:

$$\bar{x}_j = \frac{1}{N} \sum_{i=1}^{N} x_{ij}, \quad j = 1, \ldots, K.$$

Thus, the sample mean vector contains the average of the observations for each variable, and is written

$$\bar{\mathbf{x}} = \frac{1}{N} \sum_{i=1}^{N} \mathbf{x}_i = \begin{bmatrix} \bar{x}_1 \\ \vdots \\ \bar{x}_j \\ \vdots \\ \bar{x}_K \end{bmatrix}$$

**Sample Covariance** Correspondingly, the sample covariance matrix is a $N \times N$ matrix $\mathbf{V} = [\sigma_{jk}]$ with entries

$$\sigma_{jk} = \frac{1}{N-1} \sum_{i=1}^{N} (x_{ij} - \bar{x}_j)(x_{ik} - \bar{x}_k)$$

where $\sigma_{jk}$ is an estimate of the covariance between the j$^{\text{th}}$ variable and the k$^{\text{th}}$ variable of the population underlying the data.

In the form of matrix, the sample covariance is

$$\mathbf{V} = \frac{1}{N-1} \sum_{i=1}^{N} (\mathbf{x}_i - \bar{\mathbf{x}})(\mathbf{x}_i - \bar{\mathbf{x}})^{\text{T}}$$

**Exponential Smoothing** Exponential smoothing is a technique for smoothing time series data using the exponential window function. It can assign exponentially decreasing weights over the time series so that the influence of early observations vanishes as time progresses.

The raw time series is often represented by $\{x_t\}$ beginning at time $t = 0$, and the output of the exponential smoothing algorithm denoted as $\{y_t\}$, which may be regarded as a best estimate of what the next value of $x$ will be. When the sequence of observations begins at time $t = 0$, the simplest form of exponential smoothing is given by the formulas: [1]

$$y_0 = x_0$$
$$y_t = \alpha x_t + (1-\alpha) y_{t-1}, \quad t > 0$$

where $\alpha$ is the smoothing factor or called rate of decay factor, and $0 < \alpha < 1$.

By direct substitution, we find that

$$y_t = \alpha x_t + (1-\alpha) y_{t-1}$$
$$= \alpha x_t + \alpha(1-\alpha) x_{t-1} + (1-\alpha)^2 y_{t-2}$$
$$= \alpha \left[ x_t + (1-\alpha) x_{t-1} + (1-\alpha)^2 x_{t-2} \right.$$
$$\left. + (1-\alpha)^3 x_{t-3} + \cdots + (1-\alpha)^{t-1} x_1 \right] + (1-\alpha)^t y_0$$

In other words, as time passes the smoothed statistic $y_t$ becomes the weighted average of a greater and greater number of the past observations $y_{t-1}, \ldots, y_{t-}$, and the weights assigned to previous observations are proportional to the terms of the geometric progression

$$1, (1-\alpha), (1-\alpha)^2, \ldots, (1-\alpha)^n, \ldots$$

If $\alpha$ is zero, we essentially applies equal weight to each observation and if $\alpha$ is large, the influence of early observation decays quickly.

**Exponentially Weighted Sample Covariance** When we apply the exponential smoothing weight vector to our observation, the covariance matrix will be transformed into:

$$\mathbf{V} = \sum_{i=1}^{N} \omega_i (\mathbf{x}_i - \mu^*)(\mathbf{x}_i - \mu^*)^{\text{T}}$$

$$\mu^* = \sum_{i=1}^{N} w_i \mathbf{x}_i$$

where $w_i$ is the $i$-th entry of weight vector that sums to 1 (i.e. $\sum_{i=1}^{N} w_i = 1$)

The weight vector without normalization is:

$$w' = \begin{bmatrix} (1-\alpha)^T \\ \vdots \\ (1-\alpha)^2 \\ (1-\alpha)^1 \\ 1 \end{bmatrix}$$

$w$ is eventually achieved by scaling $w'$ to sum to 1.

The sample covariance estimator is often unstable for two main reasons. Firstly, it is highly sensitive to outliers in the data. Outliers can disproportionately influence the covariance estimate, leading to inaccuracies in the estimation of relationships between variables. Secondly, the sample covariance estimator can become non-singular if the number of samples $T$ is not sufficiently larger than the number of variables $N$. When the number of samples is insufficient, the covariance matrix may become singular, making it impossible to compute its inverse and consequently causing issues in portfolio optimization and other statistical analyses.

## 2.2 Ledoit-Wolf Shrinkage Covariance

As previously mentioned, one issue with the weighted sample covariance estimator is its non-singularity when the number of assets $N$ exceeds the number $T$ of available observations. This poses a significant problem in portfolio optimization. One approach to addressing this issue is by imposing some ad hoc structure on the covariance matrix, such as a factor model. However, factor models like the Barra model are often criticized for their subjectivity. Without prior knowledge about the true structure of the covariance matrix, relying on pre-specified structures can be unreliable. This lack of reliability can undermine the effectiveness of portfolio optimization techniques based on such models.

Ledoit and Wolf (2004) proposed the shrinkage estimator. Suppose we have a structured covariance matrix $F$ and sample covariance $S$ [5]. The shrunken covariance matrix is $\Sigma_{\text{shrink}}$ is a linear combination of both matrix:

$$\Sigma_{\text{shrink}} = \delta F + (1-\delta) S$$

where $\delta$ is a shrinkage constant between 0 and 1.

Ledoit and Wolf (2004) suggests a constant correlation model as the structure matrix covariance matrix $F$. It has average sample correlation of all pairs for the non-diagonal elements of the sample correlation matrix.

Each entry $f_{ij}$ of $F$ is written as

$$f_{ii} = s_{ii}, \quad f_{ij} = \bar{\rho} \sqrt{s_{ii} s_{jj}}$$

$$\bar{\rho} = \frac{2}{(N-1)N} \sum_{i=1}^{N} \sum_{j=i+1}^{N} \rho_{ij}$$

where $\rho_{ij}$ is the sample correlation between asset $i$ and $j$, and $s_{ij}$ is the sample covariance between asset $i$ and $j$.

**Choice of Shrinkage Constant** Ledoit and Wolf (2004) calibrated the shrinkage parameter $\delta$ by minimizing the Frobenius norm between the asymptotically true covariance matrix and the shrinkage estimator:

$$R(\delta) = E(L(\delta)) = E\left(\|\delta F + (1-\delta)S - \Sigma\|^2\right)$$

Under the assumption that $N$ is fixed while $T$ tends to infinity, Ledoit and Wolf (2003) proved that the optimal value $\delta^*$ asymptotically behaves like a constant over $T$. This constant, called $\kappa$, can be written as:

$$\delta^* \to \kappa = \frac{\pi - \rho}{\gamma}$$

$\pi$ denotes the sum of asymptotic variances of the entries of the sample covariance matrix scaled by $\sqrt{T}$: $\pi = \sum_{i=1}^{N} \sum_{j=1}^{N} \text{Asy Var}\left[\sqrt{T} s_{ij}\right]$.

Similarly, $\rho$ denotes the sum of asymptotic covariance of the entries of the shrinkage target with the entries of the sample covariance matrix scaled by $\sqrt{T}$: $\rho = \sum_{i=1}^{N} \sum_{j=1}^{N} \text{AsyCov}\left[\sqrt{T} f_{ij}, \sqrt{T} s_{ij}\right]$. $\gamma$ measures the mis-specification of the (population) shrinkage target: $\gamma = \sum_{i=1}^{N} \sum_{j=1}^{N} (\phi_{ij} - \sigma_{ij})^2$. Finally, we computed the empirical estimator for $\kappa$ and use it as $\delta$.

In the model implementation section, we will also incorporate cross-validation to determine the empirically optimal shrinkage constant. This approach allows us to select the most suitable shrinkage parameter based on the performance of the model on independent data subsets. By systematically evaluating the performance of different shrinkage constants through cross-validation, we can identify the one that yields the best balance between bias and variance, thus enhancing the robustness and reliability of our covariance estimation method.

### 2.3 Gerber Covariance

One common issue with many covariance matrix estimators is their reliance on product-moment statistics, such as standard deviation, which are non-robust. This becomes problematic when financial data contains numerous outliers. The presence of outliers can distort the correlation between assets in historical data series. Additionally, noise in financial data can be erroneously interpreted as meaningful information during portfolio optimization. For example, non-zero entries may appear in the correlation matrix estimator even when two assets have no meaningful correlation.

To address these issues, Gerber et al. (2021) proposed a robust co-movement measure known as the Gerber statistic. Instead of using Pearson Correlation, the Gerber statistic calculates the proportion of simultaneous co-movements in historical return samples where the amplitudes of such movements exceed a given threshold. The advantage of the Gerber statistic lies in its resilience to extremely large or small movements, making it more robust to financial time series [11].

#### 2.3.1 Gerber Covariance Matrix

Consider $k = 1, \ldots, N$ assets with $t = 1, \ldots, T$ time periods historical sample. Let $r_{tk}$ be the return of security $k$ at time $t$. For each pair of asset $(i, j)$ at each time $t$, we denote the pair of return observation at $t$ to be $(r_{ti}, r_{tj})$ as $m_{ij}(t)$, which has the following distribution:

$$m_{ij}(t) = \begin{cases} +1 & \text{if } r_{ti} \geq +H_i \text{ and } r_{tj} \geq +H_j, \\ +1 & \text{if } r_{ti} \leq -H_i \text{ and } r_{tj} \leq -H_j, \\ -1 & \text{if } r_{ti} \geq +H_i \text{ and } r_{tj} \leq -H_j, \\ -1 & \text{if } r_{ti} \leq -H_i \text{ and } r_{tj} \geq +H_j, \\ 0 & \text{otherwise.} \end{cases} \quad (1)$$

In the above equation, $H_k$ is a threshold for security $k$ that is calculated as $c \times s_k$, where $c$ is a fraction such as 0.5 (we will find optimal parameter $c$ by cross validation in section 6.2). $s_k$ is the sample standard deviation of the return of security $k$ (we will extend it to more robust measure in section 3.3.3).

The interpretation of above formulation is straightforward:

(1) $m_{ij}(t)$ is +1 if the series $i$ and $j$ simultaneously exceed the threshold in the same direction at $t$.

(2) $m_{ij}(t)$ is −1 if the series $i$ and $j$ simultaneously exceed their thresholds in opposite direction at $t$.

(3) $m_{ij}(t)$ is set to 0 if neither of two series simultaneously exceed the threshold at $t$.

The paper refers to a pair of assets that simultaneously exceed their thresholds in the same direction as concordant pair, and to one who exceed their thresholds in opposite directions as a discordant pair.

Given the above formulation, we define the Gerber statistic for a pair of assets $i$ and $j$ to be:

$$g_{ij} = \frac{\sum_{t=1}^{T} m_{ij}(t)}{\sum_{t=1}^{T} |m_{ij}(t)|} \quad (2)$$

Let $n_{ij}^c$ be the number of concordant pairs for assets $i$ and $j$, and letting $n_{ij}^d$ be the number of discordant pairs, equation (3) is equivalent to:

$$g_{ij} = \frac{n_{ij}^c - n_{ij}^d}{n_{ij}^c + n_{ij}^d} \quad (3)$$

Since the Gerber statistic calculates the number of simultaneous exceeding their thresholds, it is insensitive to extreme movements. Meanwhile, the existence of threshold also excludes small movements resulted from noise.

The matrix formulation of the Gerber statistic $G = [g_{ij}]$ is as following:

Let us define $R \in \mathbb{R}^{T \times N}$ to be the matrix of returns with entry $r_{tk}$ in its $t$-th row and $k$-th column. Let $U$ be an indicator matrix with the same size as $R$ for returns exceeding the upper threshold, having entries $u_{tj}$ such that

$$u_{tj} = \begin{cases} 1 & \text{if } r_{tj} \geq +H_j \\ 0 & \text{otherwise.} \end{cases}$$

Under above definition, the matrix of the number of samples that exceed the upper threshold is

$$N^{\text{UU}} = U^\top U$$

$n_{ij}^{\text{UU}}$ of $N^{\text{UU}}$ is the number of times when asset $i$ and $j$ exceed their upper thresholds.

Let $D$ be an indicator matrix for returns falling below the lower threshold, having entries $d_{tj}$ such that

$$d_{tj} = \begin{cases} 1 & \text{if } r_{tj} \leq -H_j \\ 0 & \text{otherwise} \end{cases}$$

The matrix of the number of samples that go below the lower threshold may be written as

$$N^{\text{DD}} = D^\top D.$$

Let $n_{ij}^{\text{DD}}$ of $N^{\text{DD}}$ be the number of times asset $i$ and $j$ goes below the lower threshold.

The matrix containing the numbers of concordant pairs is now:

$$N_{\text{CONC}} = N^{\text{UU}} + N^{\text{DD}} = U^\top U + D^\top D.$$

The matrix containing the numbers of discordant pairs is now:

$$N_{\text{DISC}} = U^\top D + D^\top U.$$

The Gerber matrix $G$ is:

$$G = (N_{\text{CONC}} - N_{\text{DISC}}) \oslash (N_{\text{CONC}} + N_{\text{DISC}}),$$

$\oslash$ is element-wise division. The corresponding Gerber covariance matrix $\Sigma_{GS}$ is then correspondingly defined as

$$\Sigma_{GS} = \text{diag}(\sigma) G \text{diag}(\sigma),$$

where $\sigma$ is a $N \times 1$ vector of sample standard deviation of historical return.

In summary, the Gerber statistic differs from other covariance matrix estimators, such as sample covariance and Ledoit-Wolf covariance, by only considering meaningful co-movements. Instead of relying on product-moment statistics that can be influenced by outliers and noise in the data, the Gerber statistic focuses solely on significant co-movements in historical return samples. This approach enhances the robustness of the covariance estimation process by filtering out irrelevant or spurious correlations, thereby providing a more accurate representation of the underlying relationships between assets.

### 2.3.2 Modification Towards Positive-Definiteness

One issue with the Gerber covariance matrix is that it is not guaranteed to be symmetric positive definite (s.p.d) when applied to real data. This poses a problem because our covariance matrix should always be positive definite to ensure that portfolio risk is greater than 0. The lack of s.p.d. property in the covariance matrix can lead to numerical instability and unreliable risk assessments in portfolio optimization. Therefore, it is crucial to address this issue when using the Gerber covariance matrix in practical applications.

The paper proposed a method of modification according to the empirical behaviour of two securities. Let $n_{ij}^{pq}$ be the number of observations for which the returns of assets $i$ and $j$ lie in regions $p$ and $q$, where region $p, q \in \{U, N, D\}$. $U$ represents upward (asset return exceed upper threshold), $D$ represents downward (asset return exceed lower threshold) and $N$ represents neutral (asset return exceed neither threshold). Two assets' co-movement therefore fall in nine regions $\{UD, UN, UU, ND, NN, NU, DD, DN, DU\}$ and we can now write $g_{ij}$ in equation (3) as:

$$g_{ij} = \frac{n_{ij}^{UU} + n_{ij}^{DD} - n_{ij}^{UD} - n_{ij}^{DU}}{n_{ij}^{UU} + n_{ij}^{DD} + n_{ij}^{UD} + n_{ij}^{DU}} \quad (4)$$

In order to have a positive semi-definite matrix, we now adjust the denominator in equation (4) to be:

$$g_{ij} = \frac{n_{ij}^{UU} + n_{ij}^{DD} - n_{ij}^{UD} - n_{ij}^{DU}}{T - n_{ij}^{NN}} \quad (5)$$

This modification essentially ignores the number of observations falling inside region $UN, DN, NU, ND$. Using the most recent 200 sample of S&P500 weekly return, we find that as the threshold parameter $c$ becomes larger, the pair of asset returns that do not simultaneously exceed the threshold observations are more concentrated only in the region $NN$, consistent with what the paper suggested.

In addition, we found it helpful to implement positive definite optimization to obtain an adjusted Gerber covariance matrix $\hat{\Sigma}$:

$$\begin{aligned} \min_{\Sigma} \quad & \frac{1}{2} \left\| \hat{\Sigma} - \Sigma_{\text{Gerber}} \right\|_F^2 \\ \text{s.t.} \quad & \Lambda_{\min}(\hat{\Sigma}) > 0 \\ & 0.25 \, \Lambda_{\max}(\hat{\Sigma}) \leq \Lambda_{\min}(\hat{\Sigma}) \end{aligned} \quad (6)$$

The first constraint ensures that we have a symmetric positive definite matrix, while the second constraint aims to control the condition number to achieve a more stable result.

We compared the Frobenius norm between the original Gerber covariance matrix and our adjusted Gerber statistics using two methods: Non-Opt, using the method suggested by the paper, and Opt, using positive definite programming. Both matrices were calculated in a rolling window manner on test data (as mentioned in Section 5). As indicated by Figure 2.1, positive definite programming (Opt) can produce a much closer approximation to our original Gerber covariance matrix. The t-value in Table ?? under the null hypothesis ($H_0$: distance=0) is larger when we use positive definite programming. Therefore, we decide to use positive definite programming to obtain a more accurate and stable Gerber covariance matrix.

Table 2.1: Statistics of Two Adjustment Methods

| Statistics | Non-Opt | Opt |
|---|---|---|
| Mean | 0.029765 | 0.017154 |
| Std | 0.012429 | 0.007107 |
| t-value | 2.394857 | 2.413559 |

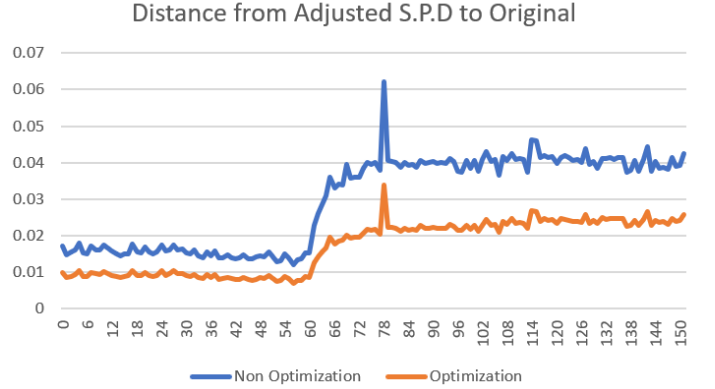

Figure 2.1: Distance

### 2.3.3 More Robust Measures for Threshold

In equation (1), $H_k$ is the threshold for security $k$ that is calculated as $c \times s_k$, where $c$ is a fraction and $s_k$ is the sample standard deviation of the return of security $k$. In this section, we want to substitute it with more robust scale estimators: median absolute deviation (MAD).

Median absolute deviation (Hampel 1974) is defined as:

$$MAD = \text{med}_i \left( \left| y_i - \text{med}_j \left( y_j \right) \right| \right)$$

where the inner median, $\text{med}_j (y_j)$, is the median of the $T$ observations and the outer median $med_i$, is the median of the $T$ absolute values of the deviations about the median. For a normal distribution, $1.4826 MAD$ can be used to estimate the standard deviation $\sigma$.

Gerber estimators with both standard deviation and MAD as thresholds will be used in portfolio for empirical analysis.

## 3 Portfolio Optimization

### 3.1 Minimum Variance Portfolio Optimization

We employed minimum variance optimization that solely relies on the estimation of covariance matrix. Kritzman, Page and Turkington (2010) points out that using historical sample to estimate asset return can be inefficient and on contrarily, we can extract more information about covariance matrix. Much literature proves that minimum variance portfolios can usually beat 1/N diversification and other weighting schemes involving estimated asset return [9].

We further augmented the minimum variance optimization with a penalty term proportional to the sum of the absolute values of the portfolio weights, which is also called $\ell_1$ norm on portfolio weights [7]. We set the parameter of $\ell_1$ regularization to be 50 basis points, which is also used as the transaction cost in consistent with Balduzzi and Lynch (1999)'s assumption [8].

We consider the following mean-variance problem with $l1$-norm transaction costs:

$$\begin{aligned} \min_{\mathbf{w}} \quad & \mathbf{w}^T \mathbf{V} \mathbf{w} + \lambda \left\| (\mathbf{w} - \mathbf{w_0}) \right\|_1 \\ \text{s.t.} \quad & \mathbf{w}^T \mathbf{1}_N = 1 \end{aligned}$$

where $\mathbf{w}$ is the portfolio weight vector. $V \in \mathbb{R}^{N \times N}$ is the estimated covariance matrix of asset returns, $\lambda \in \mathbb{R}$ is the transaction cost parameter (50 basis point), $\mathbf{w_0} \in \mathbb{R}^N$ is the weight in the beginning of the portfolio re-balance date. $\mathbf{1}_N \in \mathbb{R}^N$ is the vector of ones.

## 3.2 Nested Clustering Algorithm

Nested Clustering Optimization (NCO) is proposed by Marcos López de Prado (2019) that aims to resolve signal instability. Prado (2019) firstly identifies two kinds of instability in covariance estimator: that induced by noise and that induced by market signal itself. He argues that Ledoit Shrinkage or robust estimation method do not differentiate these two causes of instability. Instead, he suggests to only shrink the random components in the sample covariance matrix to mitigate noise instability (i.e. de-noising). Then he proposes clustering on de-noised correlation matrix to prevent signal instability [10].

### 3.2.1 Instability Caused by Noise

We first identify the instability caused by noise. Consider a matrix of independent and identically distributed random observations $X$ with $T$ observations and $N$ features (i.e. number of assets). The underlying distribution of these observations has zero mean and some variance $\sigma^2$. Then, the sample covariance matrix $V = \frac{1}{n} X'X$ has eigenvalues that asymptotically converge as $T$ goes to $\infty$ and $N$ goes to $\infty$ with $1 < \frac{T}{N} < \infty$ to the Marcenko-Pastur probability density function (Marcenko and Pastur, 1967):

$$f_\lambda(\lambda) = \begin{cases} \frac{T}{N} \frac{\sqrt{(\lambda_+ - \lambda)(\lambda - \lambda_-)}}{2\pi\lambda\sigma^2} & \text{if } \lambda \in [\lambda_-, \lambda_+] \\ 0 & \text{if } \lambda \notin [\lambda_-, \lambda_+] \end{cases}$$

The maximum expected eigenvalue is $\lambda_+ = \sigma^2 \left(1 + \sqrt{\frac{N}{T}}\right)^2$, and the minimum expected eigenvalue is $\lambda_- = \sigma^2 \left(1 - \sqrt{\frac{N}{T}}\right)^2$.

It is often assumed that eigenvalues of correlation matrix lower than $\lambda_+$ are by a chance, which we refer to 'noise' in finance, and the values higher than $\lambda_+$ are the significant common factors. We can see that covariance matrix can often contain substantial amounts of noise.

### 3.2.2 Instability Caused by Signal

Marcos López de Prado(2019) suggests that other than noise, certain covariance structures can also make the optimization problem produce unstable solutions. The easiest case is a $2 \times 2$ correlation matrix $C$:

$$C = \begin{bmatrix} 1 & \rho \\ \rho & 1 \end{bmatrix}$$

where $\rho$ is the correlation between two variables.

$|C|$ is the determinant of $C$, $|C| = 1 - \rho^2$.

By spectral decomposition on $C$, we have $CQ = Q\Lambda$ as follows, where

$$Q = \begin{bmatrix} \frac{1}{\sqrt{2}} & \frac{1}{\sqrt{2}} \\ \frac{1}{\sqrt{2}} & -\frac{1}{\sqrt{2}} \end{bmatrix}, \Lambda = \begin{bmatrix} 1+\rho & 0 \\ 0 & 1-\rho \end{bmatrix}$$

We can see that $\rho$ approaching 1 can cause $|C|$ to approach zero and the top eigenvalue to become very far away from other eigenvalues. Therefore $C^{-1}$ used in the optimal solution can be problematic. Since correlation matrix $C$ directly relates to covariance matrix $V$, we can conclude that when assets within a portfolio are highly correlated ($0 \ll |\rho| < 1$), the value of $V^{-1}$ estimator may explode and makes the optimal solution $\omega^*$ unstable. Generally speaking, one eigenvalue can only increase at the expense of the other eigenvalues given the trace of the correlation matrix $N$ (number of assets). As a result, condition number $\kappa(A) = \frac{\sigma_{\max}(A)}{\sigma_{\min}(A)}$ will be greater ($\sigma_{\max}(A)$ and $\sigma_{\min}(A)$ are maximal and minimal singular values of $A$ respectively) and yield less stable covariance estimator. Such instability is inevitable when we have a portfolio in which assets are highly correlated.

To resolve this issue, Prado (2019) proposed a method called Nested Cluster Optimization. It clusters highly-correlated assets into subsets and tries to restrict this instability into each cluster and prevent it from spreading over all assets.

### 3.2.3 De-noising

We firstly tackle the instability induced by noise. We implement Kernel Density algorithm to fit the empirical distribution of eigenvalues of our sample covariance matrix. Then we compare the theoretical distribution of Marcenko-Pastur distribution (section 3.2.1) to the empirical one so that we can determine the cut-off level $\lambda_+$ for non-random eigenvectors (separating noise-related eigenvalues from signal-related eigenvalues).

Let $\{\lambda_n\}_{n=1,\dots,N}$ be the set of all eigenvalues, ordered descending, and $i$ be the position of the eigenvalue such that $\lambda_i > \lambda_+$ and $\lambda_{i+1} \leq \lambda_+$. Then we set:

$$\lambda_j = \frac{1}{N-i} \sum_{k=i+1}^{N} \lambda_k, \ j = i+1, \dots, N$$

Given the eigenvector decomposition of covariance matrix $V$ is $VQ = Q\Lambda$, we can derive the de-noised correlation matrix $C$ as:

$$\tilde{C} = Q\tilde{\Lambda}Q'$$
$$C = \left(\text{diag}\left[\tilde{C}\right]\right)^{-1/2} \tilde{C} \left(\text{diag}\left[\tilde{C}\right]\right)^{-1/2}$$

where $\tilde{\Lambda}$ is the diagonal matrix with adjusted eigenvalues and we re-scale $\tilde{C}$ to make diagonal entries to be 1.

### 3.2.4 Clustering

Then, we tackle instability induced by signal. The Nested Clustered Optimization (NCO) employs K-means algorithm to divide the covariance matrix into $K$ groups of highly-correlated variables. The choice of optimal number of groups $K$ is based on the Z-score of sample Silhouette Coefficient (Rousseeuw, 1987). It represents the separation distance between the resulting clusters. Higher Silhouette Coefficient indicates better clustering result. The resulting clusters are $K$ subsets of our assets.

Secondly, we perform minimum variance optimization to each cluster. This can be interpreted as creating 'funds' out of our original assets and allows us to reduce the covariance matrix $V \in R^{N \times N}$ into lower dimension (number of clusters $K$). The reduced correlation matrix is closer to an identity matrix than the original correlation matrix and therefore more amenable to instability caused by signals. Finally, we performed optimization on 'funds' using the reduced covariance matrix $V_{\text{reduced}} \in R^{K \times K}$. Final weights on each original asset are dot-product of the intra-cluster weights and inter-cluster weights.

Combining de-noising in section 3.2.3 and clustering in section 3.2.4, we reach the Nested Clustering Algorithm. The advantage of NCO is that the instability only occurs within each cluster and does not propagate across clusters. Moreover, it is agnostic to what optimization method we use both intra-cluster and inter-cluster. In this paper, we implement NCO on minimum variance portfolio.

---

**Algorithm 1** Nested Clustered Optimization

**Input**: Sample covariance matrix $V$
(1) Obtain de-noised covariance matrix $\hat{V}$ and correlation matrix $\hat{C}$
(2) Cluster correlation matrix $\hat{C}$ into $K$ groups
(3) Intra-cluster opitmization within each of $K$ groups, concatenate $K$ vectors of weight into $R^{K \times N}$ weight matrix, denoted as $\Omega_{\text{intra}}$
(4) Use the intra-cluster weights to get reduced sample covariance matrix: $V_{\text{reduced}} = \Omega'_{\text{intra}} \hat{V} \Omega_{\text{intra}}$
(5) Inter-cluster optimization using $V_{\text{reduced}}$ and solve for $\Omega_{\text{inter}} \in R^K$
(6) Final weight allocation: $\Omega'_{\text{intra}} \Omega_{\text{inter}}$
**return:** final optimal weights allocated on each asset

---

## 3.3 CVaR Portfolio

Conditional Value at Risk (CVaR), introduced by Rockafellar and Uryasev (2000), is a risk measure that quantifies the amount of tail risk an asset or portfolio has [12]. To control for extreme risk of the portfolio, we apply CVaR constraints to original minimum variance optimization.

VaR estimates how much at least a portfolio might lose with a given probability or quantile. Based on this, CVaR is defined as the expectation of portfolio loss given that loss is occurring at or below the q-quantile. More specifically, CVaR is calculated by taking the weighted average of the losses above some threshold in the tail of the return distribution. Thus, CVaR works better than traditional VaR because it could deal with distribution of tails.

We tried to add to our original minimum variance portfolios one CVaR constraint ($\alpha = 0.95, \beta = 0.05$) and two CVaR constraints ($\alpha 1 = 0.95, \beta 1 = 0.05; \alpha 2 = 0.99, \beta 2 = 0.08$) respectively. The optimization is written as follow:

$$\begin{aligned} \min_{\mathbf{w},l} \quad & \mathbf{w}^T \mathbf{V} \mathbf{w} + \lambda \|\mathbf{w} - \mathbf{w_0}\|_1 \\ \text{s.t.} \quad & \mathbf{w}^T \mathbf{1}_N = 1 \\ & l + \frac{1}{1-\alpha} \sum_{\omega \in \Omega} \mathbf{P}(\omega) \max(\text{loss}(\mathbf{w},\omega) - l, 0) \le \beta \end{aligned}$$

As mentioned, $\mathbf{w}$ is the portfolio weight vector. $V \in \mathbb{R}^{N \times N}$ is the estimated covariance matrix of asset returns, $\lambda \in \mathbb{R}$ is the transaction cost parameter, $\mathbf{w_0} \in \mathbb{R}^N$ is the weight in the beginning of the portfolio re-balance date. $\mathbf{1}_N \in \mathbb{R}^N$ is the vector of ones.

In the CVaR constraint, $\omega$ is an event in the sample space $\Omega$ with probability $\mathbf{P}(w)$. We use historical return sample as the sample space and define $\mathbf{P}$ to be $\frac{1}{T}$. Thus, $loss(\mathbf{w},w)$ is sampled from historical return and defined as $R'\mathbf{w}$ ($R \in \mathbf{R}^{T \times N}$)

## 4 DATA and PORTFOLIO RE-BALANCING

We collected data of S&P500 equities over the time period January 2012 to January 2022 as well as sector information from yahoo finance. We choose S&P500 universe equities because they have high liquidity, relatively high quality and long trading period.

Our portfolios are re-balanced weekly and all data are sampled weekly so that we do not need to adjust our estimation of asset return and covariance. We employed the following procedure to benchmark performance among different covariance estimators under the context of portfolio optimization:

At the beginning of each week, the 200 weekly returns of our selected list of equities from a window are utilized to estimate our covariance matrix in the minimum variance matrix. For CVaR constraints, we will use 400 weekly samples. Then we re-balance our previous portfolios according to the optimal weight vector and hold the new portfolio for one week. At the end of the week, the realized portfolio value is computed with deducted transaction cost. If some equities are removed from the list, we will liquidate them and account for liquidation fee. We repeat this process by moving the period one week forward. This rolling-window method allows us to be more adaptive to the market.

In terms of the size of portfolio, We will use a total size of 55 individual stocks. We adopted the market capitalization regime for equity selection: choose the top 5 equities in each of 11 sector (in total 55 equities) with largest market capitalization.

We split the data into two subsets: 50% for training set and 50% for testing set for minimum variance portfolios, minimum variance portfolios using Nested Clustering Optimization, and minimum variance portfolios with CVaR constraint. Our split of data can ensure that test set includes both bull and bear market trend such as the bear market from Feb 2020 Covid-19 outburst and the subsequent bull market.

## 5 PORTFOLIO CONSTRUCTION

Our test set portfolio construction began at the end of December 2018 and will continue to be re-balanced weekly until the end of 2020. We constructed a market capitalization weighted portfolio of large-cap equities as our benchmark (refer to Figure 5.1).

To tune the parameters for the Ledoit and Gerber covariance measures, we performed 5-fold cross-validation on the training dataset. All five validation sets are disjoint from each other, and we use the Sharpe Ratio averaged over 5 folds as the evaluation criterion. The parameter corresponding to the highest Sharpe Ratio on the validation set will be used to construct portfolios on the test set. Each validation set includes 25 weekly rebalancing, holding the portfolio for approximately half a year. The transaction cost is set to be 50 basis points for both purchasing and selling.

There are three covariance matrices for which we need to tune parameters:

1.Two Gerber covariance measures (using median absolute deviation and standard deviation as the threshold, respectively) contain a parameter on the threshold ($c$ in $c \times H_k$). We tested possible values ranging from 0.3 to 1. The optimal parameter for the one using median absolute deviation is 0.4. The optimal parameter for the one using standard deviation is 0.6. 2. Ledoit covariance measure also contains a shrinkage parameter ($\delta$). We tuned from 0.1 to 1 and the optimal result is 0.4.

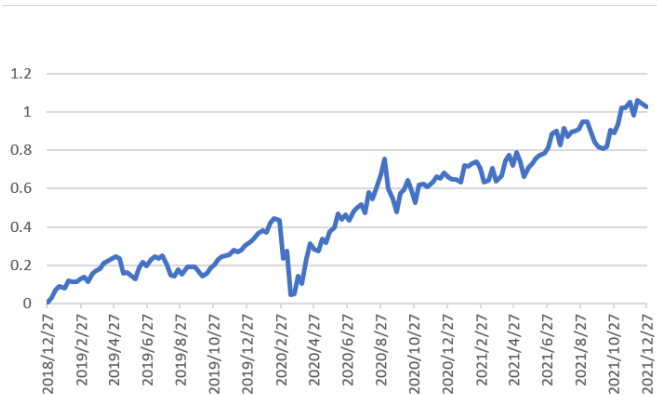

Figure 5.1: Market Cap Weighted Benchmark:2019:01-2021:12

### 5.1 Minimum Variance Performance Analysis

In this section, we first analyze performance metrics including annualized return, annualized volatility, Sharpe Ratio (with the risk-free rate set to 0), and maximum drawdown on the test data from December 2018 to December 2021. Then, we compare the differences between covariance estimators in terms of the value of correlation and their eigenvalue distribution.

The results are presented in Table 5.1. "Market" represents the capitalization-weighted portfolio. "Exp" is the minimum variance portfolio with exponentially-weighted sample covariance matrix. "Gerber_Mad" is the minimum variance portfolio with Gerber covariance matrix using median absolute deviation (MAD) as the threshold. "Gerber_Std" is the minimum variance portfolio with Gerber covariance matrix using standard deviation as the threshold. "Ledoit_Optimal" is the minimum variance portfolio using the optimal shrinkage parameter, while "Ledoit" is the one for which we tuned the best shrinkage parameter using cross-validation.

Among all portfolios, "Gerber_Mad" has the largest annual Sharpe ratio with the highest annual return and the smallest annual volatility. The excess return of the "Gerber_Mad" portfolio compared to other portfolios keeps widening starting from June 2020 (refer to Figure 5.2). Both Gerber portfolios, "Gerber_Mad" and "Gerber_Std," are in the leading position during the bull market, but "Gerber_Std" slightly underperforms "Gerber_Mad." While "Ledoit_Optimal" and "Ledoit" portfolios have smaller annual returns than the two Gerber portfolios, they can still outperform the market.

In general, each minimum variance portfolio can beat the market benchmark in terms of cumulative return if the market is in an upward trend, for example, from June 2019 to December 2019 and from April 2021 to December 2021 (refer to Figure 5.2). However, they all suffer larger losses when the market declines. The cumulative return of each portfolio does not differ much during bear markets. Due to Covid-19, the maximum drawdown all takes place during March 2020, and we can see that all portfolios suffer from an average 30% loss, which is 5% larger than the maximum drawdown of the market portfolio in this bear market (refer to Table 5.1)).

Table 5.1: MinVar Portfolio Statistics:2019:01-2021:12

|               | AnnRet | AnnVol | Drawdown | Sharpe |
|---------------|--------|--------|----------|--------|
| Market        | 31.91% | 23.16% | 25.59%   | 1.10   |
| Exp           | 31.81% | 23.12% | 31.92%   | 1.10   |
| Gerber_Mad    | 39.03% | 23.06% | 30.50%   | 1.25   |
| Gerber_Std    | 34.13% | 23.09% | 31.70%   | 1.15   |
| Ledoit_Optimal| 31.86% | 23.38% | 31.31%   | 1.12   |
| Ledoit        | 32.28% | 23.35% | 31.63%   | 1.13   |

Table 5.2: MinVar Turnover Rate

| Portfolio      | Turnover Rate |
|----------------|---------------|
| Gerber_Mad     | 7.98%         |
| Gerber_std     | 7.58%         |
| Ledoit_Optimal | 8.22%         |
| Ledoit         | 7.94%         |
| Exp            | 7.90%         |

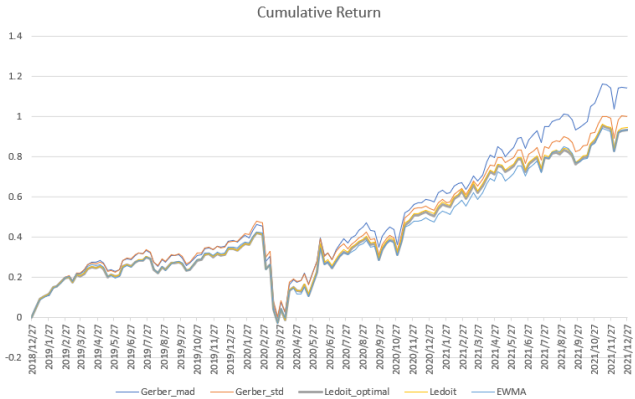

Figure 5.2: MinVar Cumulative Return

It is interesting to note that the performance of the two Ledoit covariance matrices, "Ledoit_Optimal" and "Ledoit," are similar to each other. The core difference between using the optimal shrinkage parameter and the parameter obtained through cross-validation is that the former allows us to find the optimal shrinkage matrix in an adaptive manner, while the latter is subject to back-testing limitations.

However, we found that the optimal shrinkage constant in "Ledoit_Optimal" eventually converges to the parameter obtained from cross-validation (refer to Figure 5.3). In particular, the optimal shrinkage parameter reaches the maximum possible value of 1 during March 2020. This indicates that the Ledoit shrinkage covariance matrix becomes exactly the pre-specified structure (constant correlation model) and refuses to use historical samples. This behavior is reasonable since the market becomes highly volatile during this time period and is believed to deviate from the true structure of the covariance matrix. However, it may not be optimal to perform such strong shrinkage here because we want to exploit the differences in correlation between assets for a portfolio to survive the market plunge.

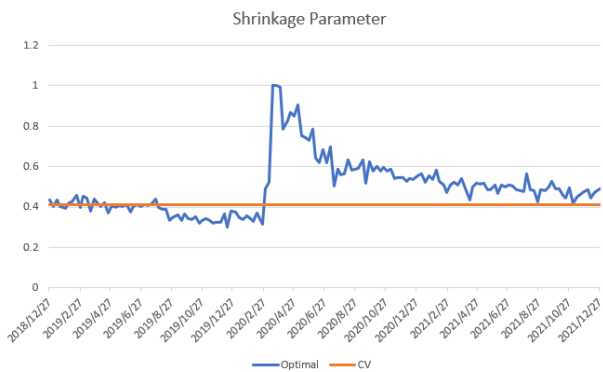

Figure 5.3: Shrinkage Parameter in Ledoit Covariance

Next, we examined whether the resulting portfolios are diversified by looking at the maximum absolute weight in each portfolio across time (refer to Figure 5.4). The weight allocations of all portfolios are relatively diversified, with the largest weight allocation being around 25%. Two portfolios, "Gerber_Mad" and "Exp," exhibit the largest changes in maximum weights over the period. On the other hand, other portfolios, including the two Ledoit portfolios and "Gerber_Std," do not allocate more than 10% weight to a single asset, and their maximum weights do not vary much across time. This is a desirable property, especially when portfolio managers need to adhere to regulatory constraints on maximum weight allocations or when regularization requires imposing constraints on maximum weight.

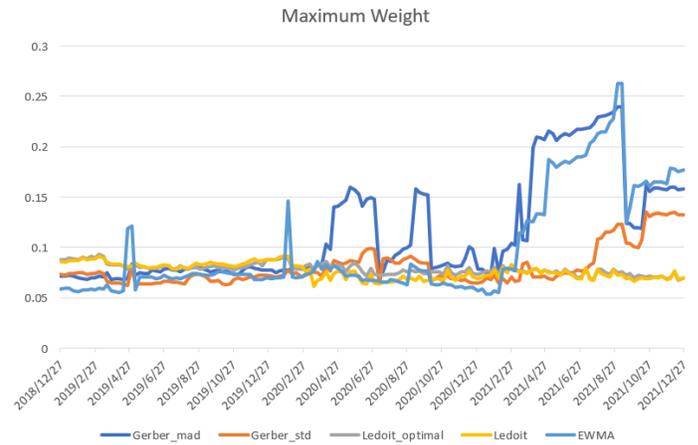

Figure 5.4: MinVar Maximum Absolute Weight

The transaction cost of all minimum variance portfolios are very similar to each other (Figure 5.5). As a proxy for transaction costs, we also calculated the average daily turnover rate and we found that all portfolios have around 8% turnover rate (Table 5.2). It could be that the $\ell_1$ regularization on weight contributes a lot to controlling the turnover rate and transaction cost.

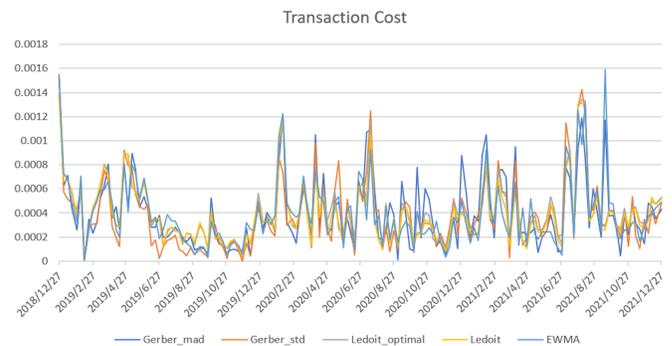

Figure 5.5: MinVar Transaction Cost

**Comparison of Covariance Matrices**  To shed more light on the introduced covariance measures, we examined all matrices using the first 200 weekly sample returns on 55 equities in the test set.

We computed two Gerber covariance matrices (Gerber_Mad and Gerber_Std), the Ledoit covariance matrix (Ledoit), and the exponentially-weighted sample covariance (Exp). Additionally, we performed de-noising method mentioned in section 3.2.3 on sample covariance to obtain the de-noised covariance matrix. This method serves as an alternative way to handle noise instability.

According to the correlation matrix heatmaps (refer to Figure 5.6), Ledoit shrinks the sample correlation matrix to the greatest extent, followed by the de-noising method.

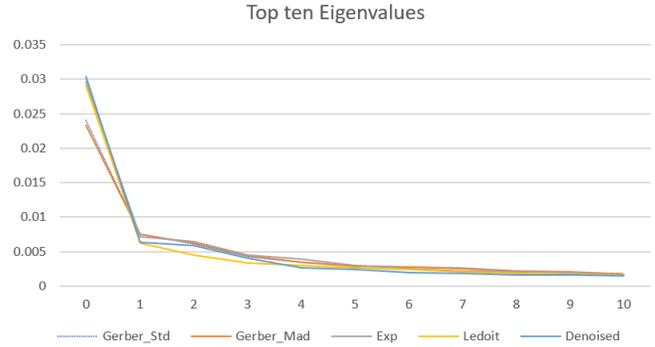

Figure 5.7: Top Ten Eigenvalues

Table 5.3: Statistics of NCO

| Statistics | NCO |
| --- | --- |
| Annual Return | 21.12% |
| Annual Volatility | 19.72% |
| Sharpe Ratio | 0.94 |
| Max Drawdown | 24.51% |

clusters are significantly separated from each other, with t-values above 2 (refer to Figure 5.8).

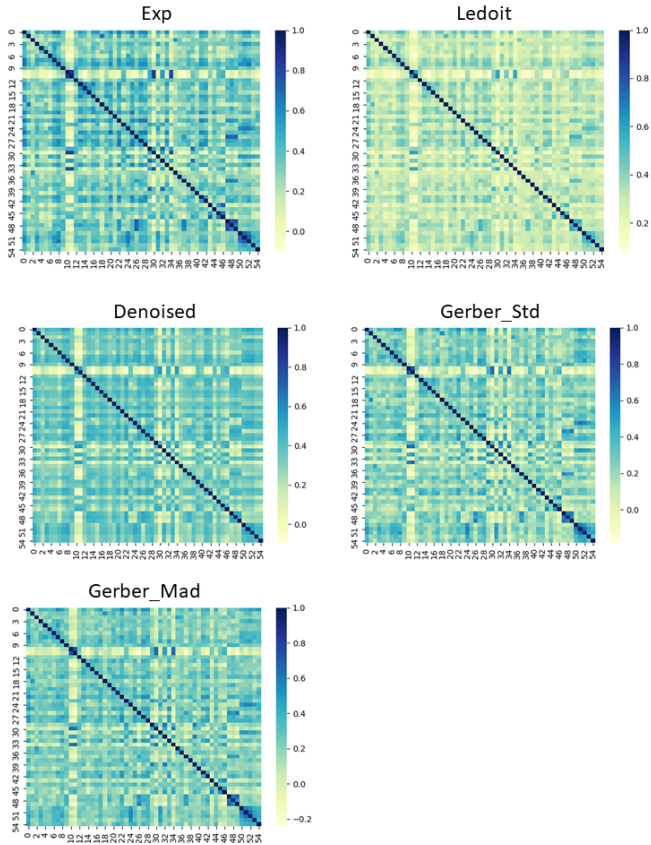

Figure 5.6: Correlation Matrix Heatmap

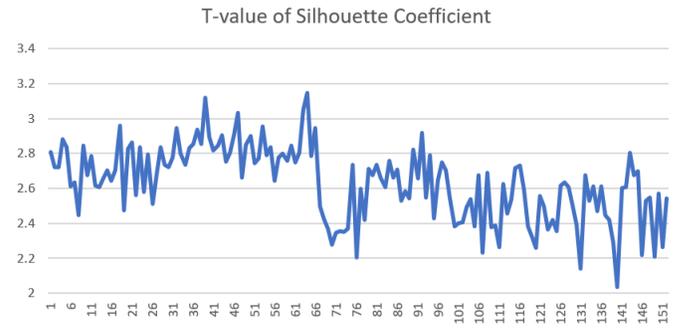

Figure 5.8: t-value of Silhouette Coefficient

Figure 5.7 presents the top ten eigenvalues of covariance matrices. We can observe that the top eigenvalue is significantly different from the other eigenvalues, which can lead to signal instability. The two Gerber covariance matrices (Gerber_Mad and Gerber_Std) are more similar to each other and have relatively smaller top eigenvalues compared to other covariance measures.

There is no significant difference between the sample covariance matrix and the Ledoit covariance matrix, as well as the de-noised covariance matrix. Although both methods aim to level eigenvalues and minimize the influence of noise, they seem to have little effect in resolving signal instability. For this purpose, we need to introduce Nested Clustering Optimization.

## 5.2 Nested Clustering Optimization Performance Analysis

Although the aforementioned covariance estimators aim to provide a more stable and accurate covariance estimation, they still suffer from signal instability, as evidenced by the presence of a top eigenvalue far from the others. This instability can lead to worse portfolio performance, particularly in bear markets.

In this section, we implemented Nested Clustering Optimization (NCO) on top of our minimum variance portfolio with the sample covariance matrix. The number of clusters across the test period ranges from 2 to 4. The t-value of the Silhouette Coefficient indicates that all

We observe that after using NCO to enhance our minimum variance portfolio, the max draw-down of minimum variance portfolio on average decreases by 5% (from 30% to 24.51%). Although we achieve less return, the performance is more robust (Figure 5.9). The annual volatility 19.72% is also the smallest among other minimum variance portfolios.

As shown in Figure 5.10, all portfolios are very diversified. A single asset will not be assigned with over 20% across test periods. Compared to the minimum variance portfolio using sample covariance matrix, the extent of diversification is improved.

The transaction cost is higher than the original minimum variance portfolios (Figure 5.11) with 25.13% turnover rate. One possible reason could be that we did not apply $\ell 1$ regularization on turnover of synthetic 'funds' in the phase of inter-cluster optimization, since no actual turnover is carried with it.

In conclusion, Nested Clustering Algorithm can reduce volatility and portfolio draw-down and gives a more diversified portfolio.

## 5.3 CVaR Portfolio Performance Analysis

In this section, we consider another extension on minimum variance portfolio by adding Conditional Value-at-Risk constraint. While Nested Cluster Optimization focuses on improving the covariance measure, CVaR introduces a new and coherent risk measure to account for extreme risk. It can be helpful when we are faced with extreme market condition. We constructed the minimum variance portfolio with one CVaR constraint and two CVaR constraints.

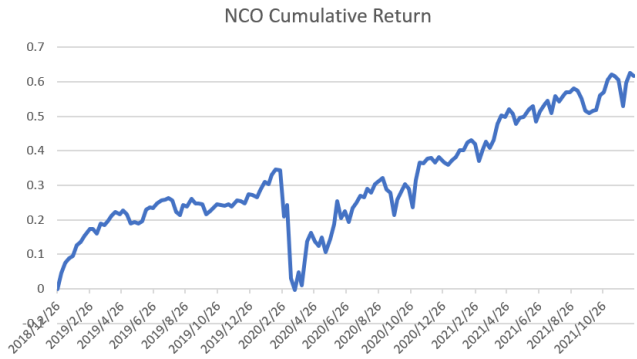

Figure 5.9: NCO Cumulative Return

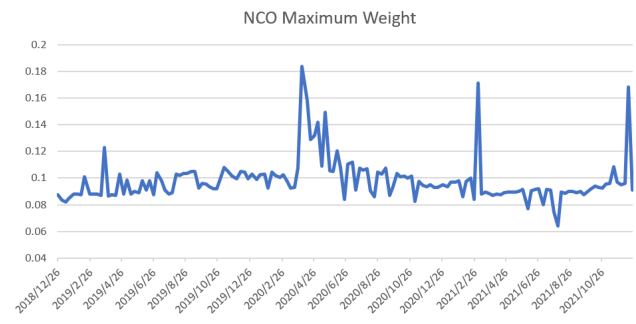

Figure 5.10: NCO Maximum Absolute Weight

**Scenario 1** ($\alpha = 95\%, \beta = 5\%$) The performance of minimum variance portfolio improved when we require weekly CVaR at 95% level to be smaller than 5%.

As shown in table 5.4, all portfolio max draw-down on average decrease by 9% during March 2020, the period of pandemic. CVaR constraint minimum variance portfolios also have on average 4% lower max draw-down than the market (i.e. market capitalization weighted portfolio). In terms of annual volatility, adding CVaR constraint makes Ledoit_Optimal and Ledoit portfolio have 3% less volatility. In terms of Sharpe ratio, all portfolios have improved by more than 0.2. For instance, Sharpe ratio of Gerber_Mad portfolio increases from 1.25 to 1.45. To further demonstrate the effect of CVaR constraint, we calculated Sortino ratio. It replaces the volatility in Sharpe ratio with semi-volatility (volatility of negative return). Given the fact that Sortino ratio of all portfolios are larger than Sharpe ratio, we can infer that the the loss of our portfolio is less volatile than the gain.

Figure 5.12 shows the maximum weight allocated to each asset in the portfolio. Portfolios using Gerber Mad covariance (Gerber_Mad) and exponentially weighted covariance (Exp) have the most variable weight allocation. Interestingly, all portfolios choose to put more than 30% weight on some asset around March 2020. After investigating into weight allocation (Figure 5.13), we found that they all tend to long The Clorox (NYSE: CLX) on March 20, 2020. It is likely that longing CLX allows the portfolios' CVaR to be smaller than the threshold $\beta = 5\%$.

Table 5.4: CVaR Portfolio Statistics Scenario1: 2019:01-2021:01

| alpha=95%<br>beta=5% | AnnRet | AnnVol | Drawdown | Sharpe | Sortino |
|---|---|---|---|---|---|
| Exp | 31.85% | 19.85% | 21.61% | 1.25 | 1.37 |
| Gerber_Std | 36.87% | 20.28% | 22.78% | 1.35 | 1.44 |
| Gerber_Mad | 40.45% | 20.01% | 22.90% | 1.45 | 1.54 |
| Ledoit_Optimal | 34.48% | 20.57% | 21.64% | 1.27 | 1.37 |
| Ledoit | 34.44% | 20.25% | 21.92% | 1.29 | 1.35 |

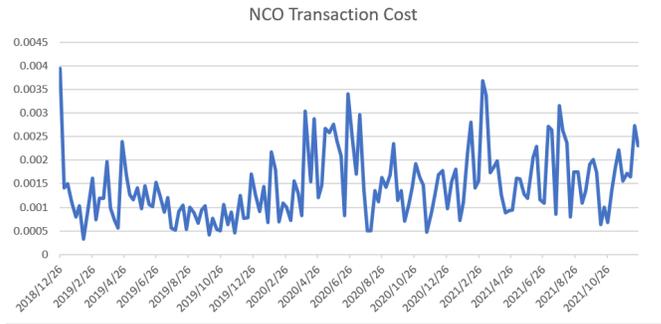

Figure 5.11: NCO Transaction Cost

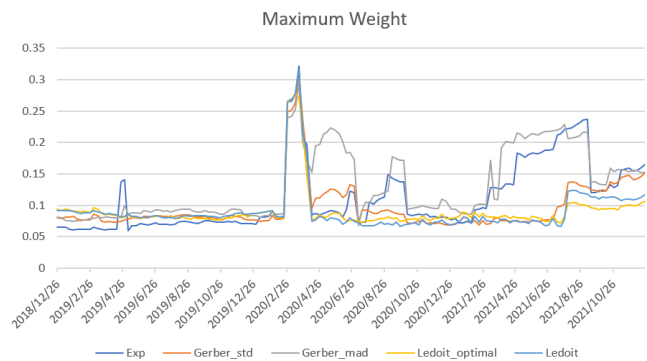

Figure 5.12: CVaR Scenario1 Maximum Absolute Weight

The pie chart (Figure 5.13) gives an example of weight allocation of our 1-CVaR Min-Var portfolio using exponentially weighted covariance on day March 20, 2020. We can see that the stock CLX takes the largest proportion.

Figure 5.14 shows the cumulative return for all five portfolios with different covariance matrix measure hes. We can see that all the cumulative returns are in a smoothing increasing trend despite a sharp downturn in the first quarter of 2020. Moreover, similar to the results of minimum variance portfolios without CVaR constraint, 1-CVaR portfolio using Gerber Mad covariance (Gerber_Mad) have the highest cumulative return than other portfolios over most periods.

The transaction cost of all portfolios are very similar and comparably small (Figure 5.15). It is however, slightly higher than original minimum variance portfolio.

**Scenario 2** ($\alpha_1 = 95\%, \beta_1 = 5\%$ $\alpha_2 = 99\%, \beta_2 = 8\%$) In addition to requiring weekly CVaR at 95% level to be smaller than 5%, we further constraining CVaR at 99% level to be smaller than 8%. We found that adding another CVaR constraint produces portfolios with lower risk.

As shown in table 5.5, the most significant improvement is max draw-down: 2-CVaR constraint minimum variance portfolios have more than 2% lower max draw-down than the 1-CVaR constraint portfolio. The reduction in annual volatility is, however, less noticeable. Adding one more CVaR constraint reduce the volatility of all portfolios by 1%. In terms of Sharpe Ratio, all portfolios have a slightly lower Sharpe Ratio compared to 1-CVaR constraint portfolios. All portfolios Sortino ratio are larger than Sharpe ratio and the gaps become wider compared to 1-CVaR portfolios. For example, Gerber_Std in 2-CVaR has larger Sortino Ratio despite a similar Sharpe ratio. Sortino ratio of Ledoit optimal is 0.06 higher than Sharpe ratio in 1-CVaR and 0.1 higher than Sharpe ratio in 2-CVaR. These results are consistent with the fact that portfolios with more CVaR constraints are more risk-adverse.

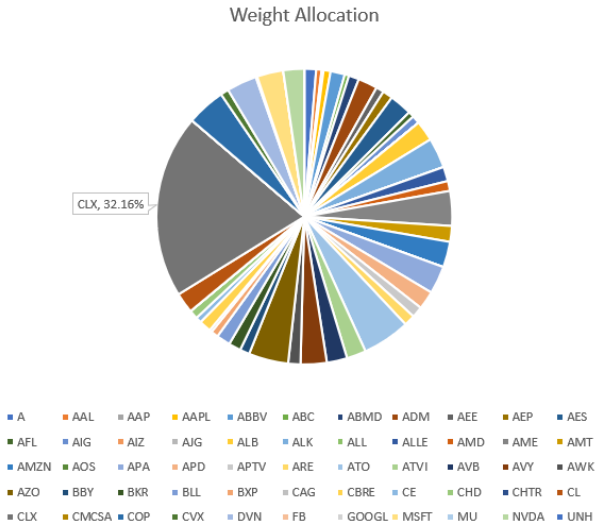

Figure 5.13: CVaR Scenario1 Weight Allocation:2020:03:20. Using portfolio with exponentially-weighted covariance as an example.

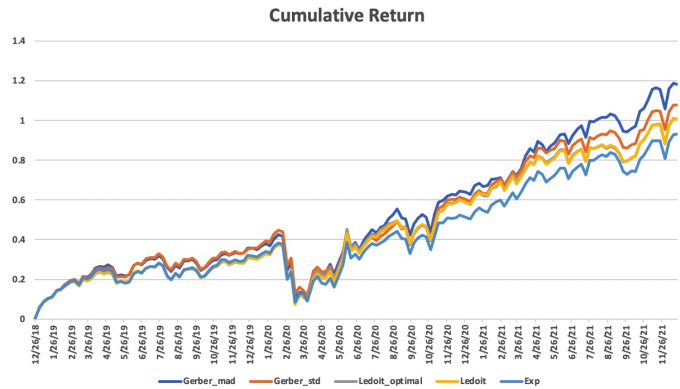

Figure 5.14: CVaR Scenario1 Cumulative Return

Overall, using two CVaR constraints can better control for risks, but the improvement in performance is less obvious compared to the improvement from 1-CVaR portfolio. Therefore, we can conclude that using one CVaR constraint is usually adequate for the purpose of portfolio risk management.

Figure 5.16 shows the cumulative return of each portfolio. Similar as 1-CVaR-portfolios, Gerber_Mad is still in leading position, but the difference between Gerber_Mad and Gerber_Std becomes smaller. We can see that portfolios have milder draw-down during March 2020 compared to the original minimum variance portfolios.

Figure 5.17 shows the maximum weight allocation in each portfolio. Compared to portfolio with only one CVaR constraint, the maximum weight assigned to each asset across time increases from 30% to more than 35%. While all portfolios still agree on putting the most weight on the Clorox (NYSE:CLX) during March 2020, there are some additional periods where portfolios have the heaviest weight on same asset: August 3,2021, November 24, 2021 and December 16,2021. For all portfolios, the most concentrated asset on August 3,2021 is Google (NYSE:GOOGL) and the most concentrated asset on November 24, 2021 and December 16,2021 is Autozone (NYSE:AZO). It is possible that with one more CVaR constraint, portfolios become more risk-adverse and hence all of them tend to prefer some safe assets. This leads to a less diversified portfolio compared to the original minimum variance portfolio.

The transaction costs of all 2-CVaR portfolios have the most consistent pattern compared to each other than all previously mentioned portfolios. They also have high transaction cost. The turnover rate increase by almost 8% from original minimum variance portfolio (Table 5.6). We note that transaction costs peaked on August 3,2021, November 24,2021 and December 16,2021. This coincides with our previous find on maximum weight allocation. It further stresses on the fact that with two CVaR constraints, portfolios tend to invest on some common safe assets despite the cost penalty.

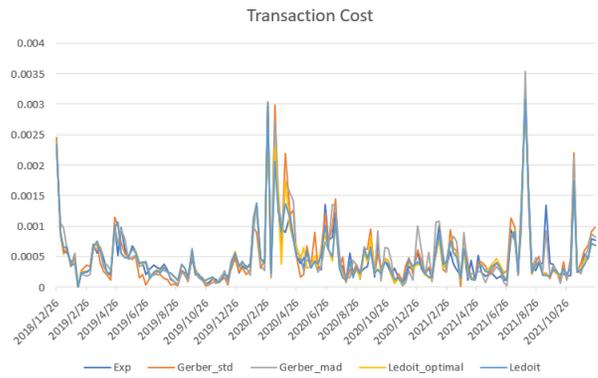

Figure 5.15: CVaR Scenario1 Transaction Cost

Table 5.5: CVaR Portfolio Statistics Scenario2: 2019:01-2021:01

| alpha=95% beta=5% alpha=99% beta=8% | AnnRet | AnnVol | Drawdown | Sharpe | Sortino |
| --- | --- | --- | --- | --- | --- |
| Exp | 28.13% | 18.51% | 19.62% | 1.21 | 1.35 |
| Gerber_Std | 33.37% | 18.85% | 20.68% | 1.34 | 1.47 |
| Gerber_Mad | 35.90% | 18.75% | 21.19% | 1.42 | 1.55 |
| Ledoit_Optimal | 30.37% | 19.26% | 19.86% | 1.24 | 1.34 |
| Ledoit | 29.64% | 18.85% | 20.09% | 1.24 | 1.34 |

Table 5.6: CVaR Scenario2 Turnover Rate

| Portfolio | Turnover Rate |
| --- | --- |
| Exp | 15.05% |
| Gerber_Std | 15.70% |
| Gerber_Mad | 16.50% |
| Ledoit_Optimal | 13.87% |
| Ledoit | 13.96% |

## 6 Conclusion

Firstly, we explored different methods to obtain covariance estimators less prone to financial noise and constructed regularized minimum variance portfolios using 55 S&P 500 Large Cap equities. We compared the performance of the Exponentially Weighted Sample Covariance, Ledoit Shrinkage Covariance, and Robust Gerber Covariance. Out-of-sample performance shows that all three covariance measures outperform the market capitalization-weighted benchmark in terms of the Sharpe ratio and achieve a return premium during bull markets. Among them, the Gerber covariance using MAD is the leading performer. The portfolios are diversified and incur minimal transaction costs. However, they all exhibit higher volatility and greater losses during bear periods compared to the market benchmark.

Next, we implemented Nested Cluster Optimization (NCO) to mitigate signal instability in the covariance estimator. The resulting minimum variance portfolio with NCO decreases the maximum drawdown

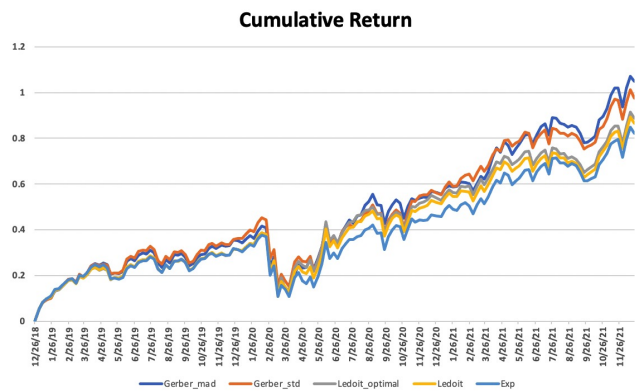

Figure 5.16: CVaR Scenario2 Cumulative Return

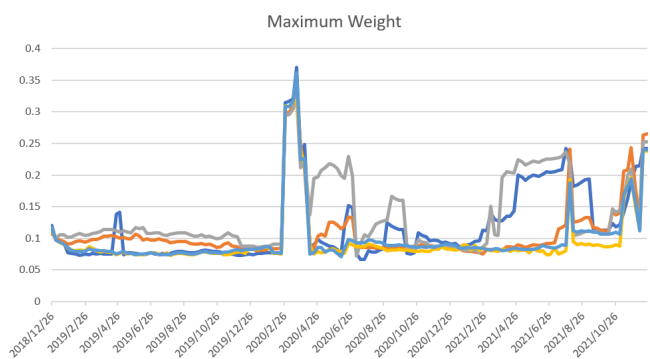

Figure 5.17: CVaR Scenario2 Maximum Weight Absolute Allocation

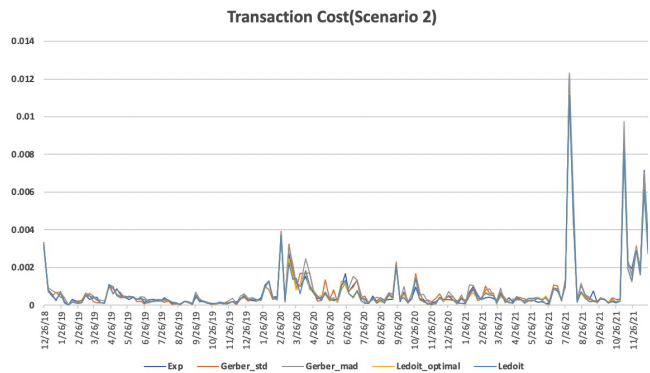

Figure 5.18: CVaR Scenario2 Transaction Cost

by 5%, despite having a more conservative return and higher transaction costs.

Finally, the introduction of constraints on Conditional Value-at-Risk (CVaR) significantly improves portfolio performance. Specifically, the minimum variance portfolio with one CVaR constraint has a 4% smaller maximum drawdown compared to the market portfolio. Adding a second CVaR constraint further reduces the maximum drawdown, but the improvement is less noticeable. Moreover, adding more CVaR constraints can lead to a more risk-averse weight allocation, resulting in a less diversified portfolio. Therefore, one CVaR constraint with a discrete choice of threshold should be adequate for managing extreme risk.